\documentclass{optica-article}

\journal{opticajournal} 

\articletype{Methodology}

\usepackage{lineno}

\usepackage{float}

\begin{document}

\title{Laser-pump-resistive-probe technique to study nanosecond-scale relaxation processes}

\author{Mark I. Blumenau,\authormark{1, 2, *} Aleksander Yu. Kuntsevich,\authormark{1}}

\address{\authormark{1}P.N. Lebedev Physical Institute of the Russian Academy of Sciences, Leninsky prospekt 53, Moscow, 119991, Russia\\
\authormark{2}HSE University, 20 Myasnitskaya ulitsa, Moscow 101000, Russia\\
}

\email{\authormark{*}mblumenau@hse.ru} 


\begin{abstract*} 
Standard optical pump-probe methods analyze a system's temporal response to a laser pulse within sub-femtoseconds to several nanoseconds, constrained by the optical delay line's length. While resistance is a sensitive detector in various fields, its measurements are typically slow (>microseconds) due to stabilization requirements. We suggest here a
time-resolved pump-probe technique which combines an optical pump pulse and a rectangular electrical probe pulse through the sample, measuring transmission in a 50 Ohm matched circuit with a digital oscilloscope. This allows electrically-driven delays from nanoseconds to seconds. Demonstrations include studying heat-induced changes in a thin amorphous VO$_x$ film and carrier relaxation in a CdS photoresistor, showcasing potential applications in heat transfer, biochemical reactions, and gradual electronic transformations.
\end{abstract*}

\section{\label{sec:introduction}Introduction}

The optical pump-probe technique is one of the major tools for exploration of variuos phenomena in time domain. It 
employs two laser pulses: The first ("Pump") excites the system, and the second ("Probe") measures the response after a delay. The delay time may be as small as hundreds of attoseconds \cite{art:Lara_attosecond} and frequencies could extend from UV to THz range\cite{art:Wang:07, art:Hoffmann:09}. While the optical pump-probe has has no alternative at relatively short timescales, it is rarely used for longer duration phenomena\cite{art:Younesi:22, art:Kim2018}. At the same time, in the range above 1 ns, many processes may occur in the solid state, such as temperature redistribution, mechanical micro-motion, electron relaxation on impurity states in semiconductors\cite{bk:Ryvkin2012}, or macroscopic order parameter relaxation\cite{art:Saol_macroscopic, art:He_microsecond}. The limited usage of the optical pump-probe for studies of these phenomena is due to two main reasons: (i) it is technically hard to measure at delays longer than several ns, due to the length of the delay line (1 ns corresponds to 30 cm); (ii) reflectivity, transmission or higher harmonic generation might be insensitive to the relevant physics.

Conductivity reflects the immediate vicinity of the Fermi energy and might, thus, be sensitive to very subtle changes in temperature, deformation, energy spectrum, etc. 
A combination of optical excitation and time-resolved resistive measurements seems to be a powerful tool for exploring new intermediate time-scale phenomena. 
However, resistance measurements are usually slow due to current/voltage ratio stabilization and processing, including digitalization. Conversely, an impedance-matched voltage pulse can be rapidly measured using an oscilloscope.

An idea of the technique combining an optical pump and a pulsed electric probe is straightforward. We develop a Pump-probe-like technique that would allow us to measure the relaxation of the arbitrary resistance, higher than 50 Ohms, on a time scale from nanoseconds to seconds. 

The proposed technique shares commonalities with earlier experimental studies. For example, an approach with DC bias current and voltage and a sensing resistor is sometimes used,  e. g. in Refs.~\cite{CrunteanuStanescu2020,  Crunteanu}.
A technique with an electrically pumped time-resolved Kerr rotation to study the dynamics of the spin Hall effect was implemented by Stern et al.\cite{art:Stern_hallnano}. To achieve 50 Ohm matching they adjusted carefully the sample's geometry. An all-electrical pump-probe technique was demonstrated by Dirisaglik et al. \cite{art:Dirisaglik_electrpump}. They used various measurement waveforms to study a Ge$_2$Sb$_2$Te$_5$ phase change memory device. Time-resolved carrier dynamics with electrical pumping and conductivity measurements were performed in Refs.~\cite{art:Lutz1993, art:Zhitenev1993, art:Ramamoorthy2015, art:Vasile2006, art:Naser_50ohmsemic}. In these studies
50 Ohm impedance-matched circuits and pulse widths from a couple of nanoseconds to a few hundred nanoseconds were used. In Ref.~\cite{nano_Budden}  photo-induced superconductivity in K$_{3}$C$_{60}$ was studied using optical-pump-resistive-probe technique. Despite some similarities with the technique described in the present paper,  Ref.~\cite{nano_Budden} considers low resistance samples (<50 Ohms), and small relaxation times (< several microseconds).

As we show here, arbitrarily long delay time makes the synchronisation of optical and electrical parts tricky. We provide details on syncing the laser system and the pulse generator, and handling resistances much larger than 50 Ohms. We believe this paper is of particular interest for experimental studies of novel switchable materials \cite{art:Zhang2016, art:Vaskivskyi2015, art:Lysenko2007, Badri}. 

\section{\label{sec:methodology}Method}
\subsection{\label{subsec:measuremnt}Description of the experimental technique}

\begin{figure}[H]
\includegraphics[width=0.99\columnwidth]{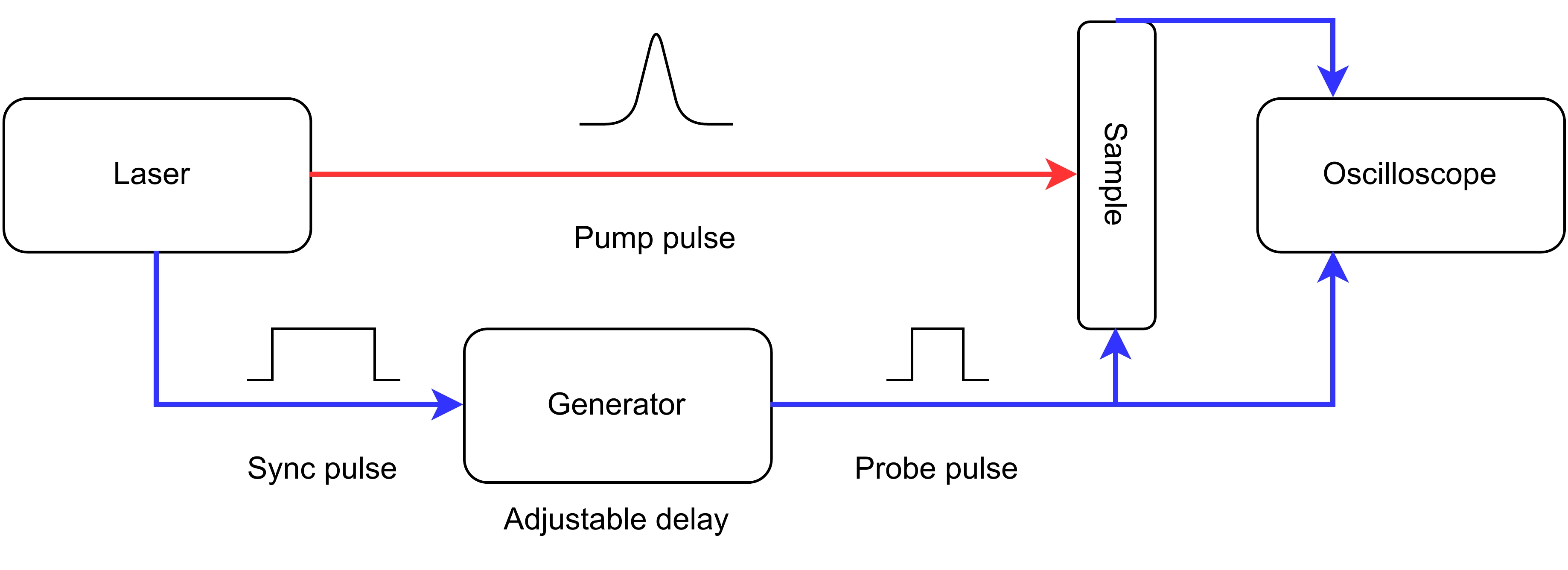}
\caption{\label{fig:exp_simple}Simplified schematics of the experimental setup. Laser pulses are shown in red, and electrical pulses are shown in blue.}
\end{figure}

The main idea of the laser-pump-resistive-probe method is simple. One needs three tools: a laser, a pulse generator, and a digital oscilloscope. A laser system outputs an optical pump pulse as well as an electrical sync pulse. The laser beam is focused on the sample with the help of mirrors and a microscope lens (if needed). The electrical sync pulse is used to trigger a pulse generator, which outputs the probe pulse after an adjustable delay. Using an oscilloscope one can measure the voltage and calculate the resistance later. A simplified schematics is shown in Fig. \ref{fig:exp_simple}. Although the idea is simple, the practical realization is complicated due to impedance matching and unwanted time delays (e.g. pulse generator input-to-output delay). 

Using short-pulsed resistivity measurements instead of DC bias current and voltage readout is an important feature, that gives two major advantages. First, the delay time is practically unlimited. Oscilloscopes cannot record data with a high sample rate (several GSa/s) for several seconds due to memory limitations. Second, the pulse technique reduces the heating of the sample. For example, a typical 10-25 ns 3 V pulse in 1 kOhm sample with a 52.5 Ohm matching resistor releases $\sim$10$^{-11}$ Joules of heat, that is less than laser pulse ($\sim$10$^{-4}$ to $\sim$10$^{-10}$ Joules, depending on the system). The resistivity measurement performed, for example, for 3 microseconds could be higher than the typical Ti:Sa femtosecond laser-pulse caused heating, while a 10 ms measurement is surely higher. 

\subsection{Experimental setup}

\subsubsection{Instruments}
We use a Light Conversion PHAROS laser system with a wavelength of 1030 nm and a minimal pulse duration of $\sim$250 fs for ``Pump'' excitation. The laser system runs at a repetition rate of 25 kHz and is accompanied by a built-in pulse picker. The pulse picker allows to further reduce the frequency of laser pulses. 

One of two options is used for probe pulse synthesis: (i) G5-78, a fast analog pulse generator, that allows to achieve pulses with widths down to 1 ns; (ii) Tektronix AFG3152C, with a digitally controlled trigger to output delay (from hundreds ps to several seconds) and pulses with widths down to 5 ns. It is an arbitrary function generator, however, we use only the rectangular pulse mode. The measured jitter is around 50 ps for the G5-78 and around 75 ps for the Tektronix AFG3152C.

To collect the voltage data we use a Keysight DSO-S104A oscilloscope with 1 GHz bandwidth, a 10-bit analog-to-digital (ADC) converter, and up to 20 GSa/s acquisition rate in 2-channel mode or 10 GSa/s in 4-channel mode.

\subsubsection{\label{synchproper}Synchronization}
Note that synchronization is highly dependent on the laser system used, and this part may differ significantly.
A Q-switched laser could be triggered from the same pulse generator. However, such lasers have limitations either in pulse duration (above several ns) or in pulse energy. For most ps and fs lasers, one requires synchronization with the internal timing module of the laser, which might be problematic, especially for delays close to zero and "Pump" repetition rates below 1 Hz. 

\begin{figure}[H]
\includegraphics[width=0.99\columnwidth]{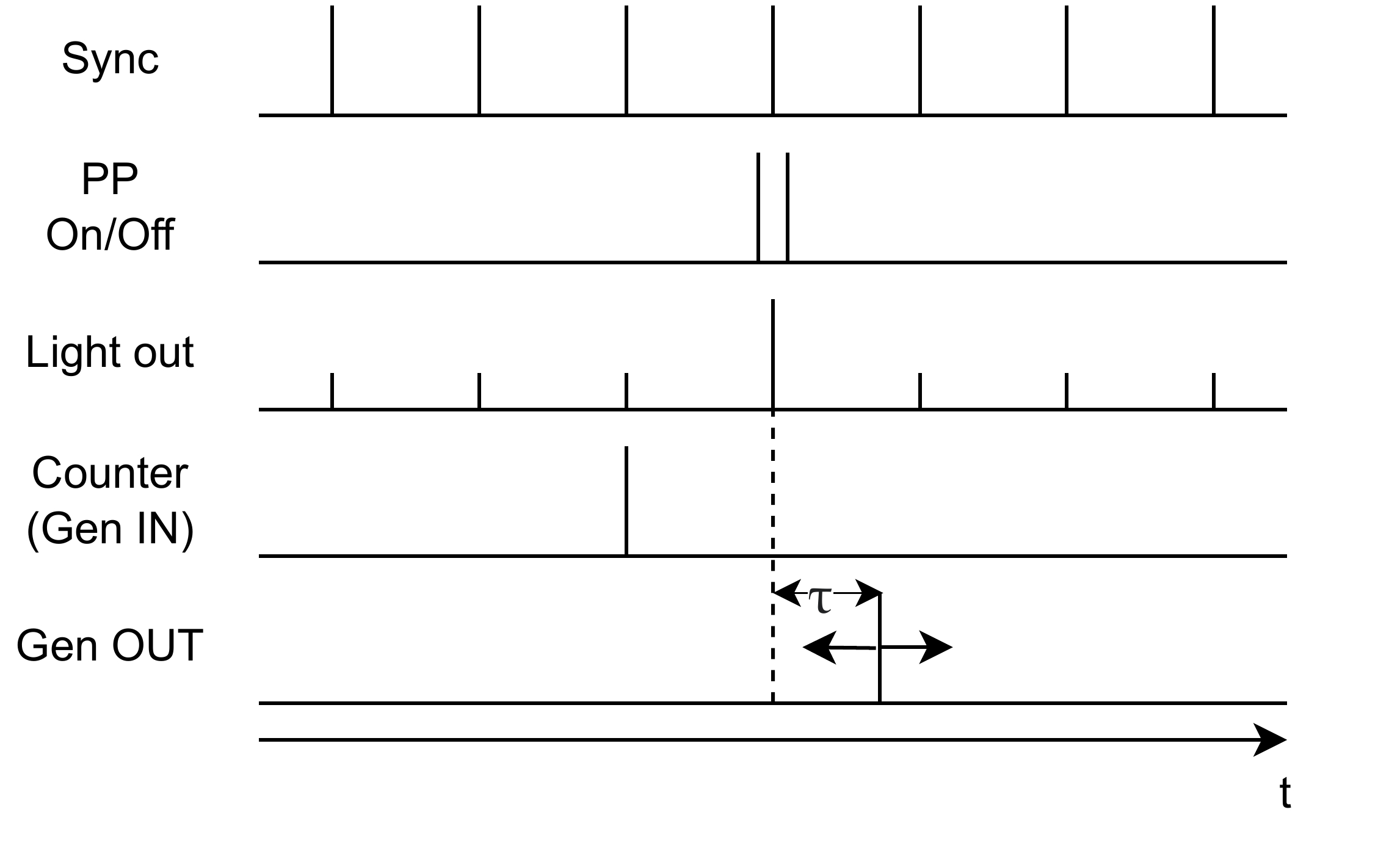}
\caption{\label{fig:pulses} Schematically shown pulse trains in our setup. Sync is the synchronization pulse train from the Pharos laser system TEM. PP is the pulse picker signal, Gen OUT is the pulse generator's output. Light out is the Pharos built-in photodiode signal.}
\end{figure}

We provide details of the synchronization in our setup below.  The relevant pulse trains are shown schematically in Fig. \ref{fig:pulses}. The laser oscillator (OSC) is running with a repetition rate of 76 MHz, inaccessible for user. The regenerative amplifier (RA) is operated at 25 kHz. Thus, a laser pulse is emitted every 40 $\mu$s. Using the PHAROS timing electronics module (TEM),  we can obtain a 25 kHz Sync pulse train of the RA.  With a photodiode on the output we can control the light output timing (Light out pulse train in Fig. \ref{fig:pulses}). The time between laser pulses is configurable with a built-in pulse picker (PP) and is chosen to allow the system under study to relax.

The PHAROS TEM outputs a Pulse Picker On and Off (PPOn/Off) signal every time a pulse is picked (1 PPOn and PPOff pulse per every picked pulse). The PPOn signal is emitted about a hundred ns before the light pulse and could be used to take the extra time delays into account. Time delays mainly consist of input-to-output delay of digital pulse generators. In our case, for Tektronix AFG3152C we measured a minimal input-to-output delay of about 300 ns. Thus, the PPOn pulse cannot be used for synchronization. If the PP outputs a light pulse, for example, every two seconds, time points 0 ns and 2 s should be equivalent. At high delays the pulse generator has only a 100-microsecond resolution (5 digits), which is much higher than 1 ns and cannot be used for studying fast dynamics. Such input-to-output delays and limited resolutions are common for most digital pulse generators.

To solve the problem, a homemade circuit to pretrigger the pulse generator is used. It consists of 6 cascaded 74HC192 presettable counters (cascading is done according to the datasheet) and an SN74AC04DR inverter. This cascade outputs an electric pulse one Sync pulse (40 $\mu$s) before the picked pulse. For a pulse picked once every two seconds, the cascade will output on the 49999th Sync pulse (because $2 \text{ seconds} \cdot 25 kHz = 50000$). The PPOff pulse, which is output by the TEM after the light pulse, is fed through the inverter to the parallel load pin of all counters to reset them and load values from data pins. PPOff comes before the next Sync pulse. The cycle repeats. The output of the cascade is used to trigger the pulse generator. The Tektronix AFG3152C generator output pulse will match the light pulse arrival if the delay $\tau$ is set to about 40 $\mu$s (depending on the length of cables compared to the light path length).The 5-digit resolution of the pulse generator is sufficient for adjusting the delay on the nanosecond scale.

For the 2 second example, the picked laser pulse should be out 40 $\mu$s after the 49999th sync pulse. However, because the whole laser is synchronized with a 76 MHz crystal, the laser pulse can arbitrarily shift by about 13 ns. Thus, to ensure proper synchronization, the third channel of the oscilloscope is used to record the PHAROS built-in output photodiode signal.

The PP never suppresses the laser pulse completely. This is shown in the schematic representation (Fig. \ref{fig:pulses}): the leaking laser pulses are drawn to be smaller. One should either check that such residual pulses do not affect the system under study or suppress them with a nonlinear filter.

\subsubsection{Electrical circuit}
\begin{figure}[H]
\includegraphics[width=0.8\columnwidth]{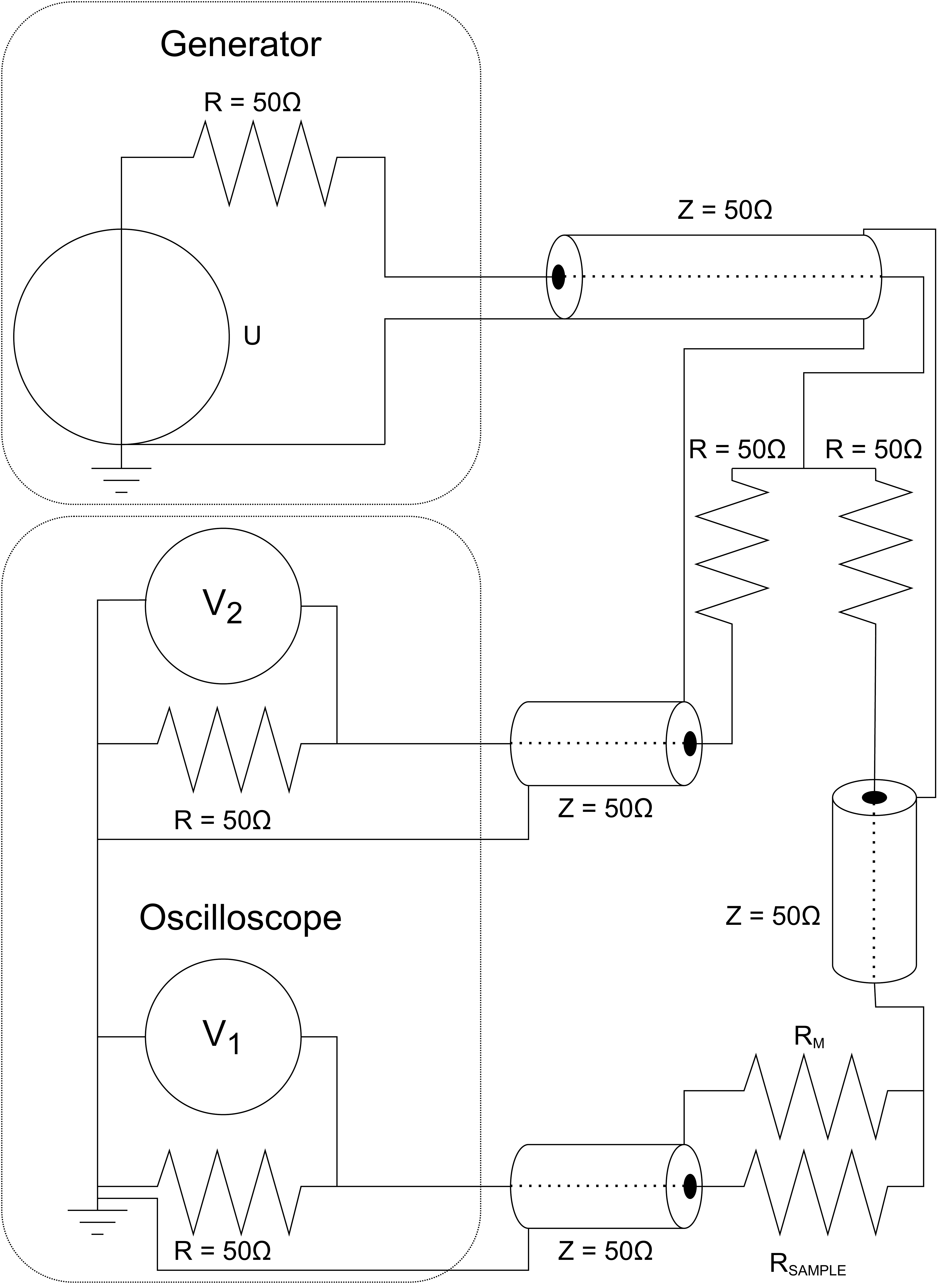}
\caption{\label{fig:electr_circuit}Schematics of the equivalent electrical circuit. The coaxial cables are shown as cylinders. Signals from the pulse generator are split with two 50 Ohm resistors. $R_M$ is selected for 50 Ohm impedance matching and $R_{SAMPLE}$ represents the sample.}
\end{figure}

The equivalent electrical scheme of the measuring circuit is shown in Fig. \ref{fig:electr_circuit}. The pulse from the pulse generator is divided: one part goes through the sample and the matching resistor to the 1st channel of the oscilloscope, and the other part goes to the 2nd channel of the oscilloscope. 50 Ohm coaxial cables are used for connections between the instruments and the sample.

In a perfectly matched circuit, the resistance of the sample should be expressed by the formula (derivation in Supplement 1):
\begin{equation}
    R_{\text{SAMPLE}}=(50 \text{ Ohm})\left(\frac{2 V_2 / V_1-1}{1+50 / R_\text{m}}-1\right),
    \label{eq:resistance}
\end{equation}

where $R_{\text{SAMPLE}}$ is the sample resistance, $R_\text{M}$ is the resistance of the impedance matching resistor, $V_1$ and $V_2$ are the voltages on the 1st and 2nd channels of the oscilloscope, respectively. Note that sample resistance should not be too high, otherwise, $V_1$ will be immeasurably small. Low resistance (< 50 Ohm) samples require a different schematic, similar to one in Ref.~\cite{nano_Budden}.

\subsubsection{\label{sssec:post}Post-processing}

\begin{figure}[H]
\includegraphics[width=0.99\columnwidth]{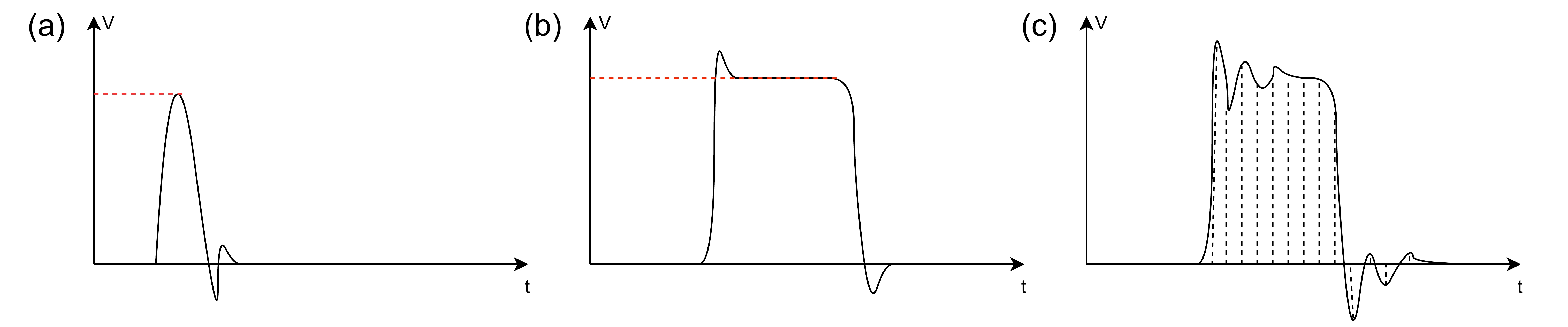}
\caption{\label{fig:three_types}Schematic representation of three methods to determine $V_1$ and $V_2$ for the sample's resistance. (a) maximum value of the oscillogram; (b) mean value of a stabilized signal; (c) numerical integration of the oscillogram.}
\end{figure}

Probe pulses in the described setup can be as short as 1 ns. However, due to unavoidable capacitive spikes in the beginning and at the end of rectangular pulses, such short widths can be used only when the circuit impedance is very close to 50 Ohm. Usually it is needed to increase the width of the probe pulse.
Similar problems were described by the other researchers, see Refs.~\cite{art:Vasile2006, art:Naser_50ohmsemic}.
Depending on the system under study (mostly on its electrical capacitance) one of three methods can be used to numerically determine $V_1$ and $V_2$ from the raw oscilloscope data:
\begin{enumerate}
    \item Maximum value on the oscillogram (for probe pulses shorter than 20 ns), schematically shown in Fig. \ref{fig:three_types}(a).
   \item Mean value of the high level (for probe pulses longer than 20 ns or with short lead and trail), Fig. \ref{fig:three_types}(b).
    \item Numerical integration of the signal part of the oscillogram (if the pulse shape is smeared by transient processes), Fig. \ref{fig:three_types}(c).
\end{enumerate}

The data is collected over different delays using a Python~\cite{10.5555/1593511} script. Several oscillograms (usually 10-20) are acquired and averaged for each delay value. The code for data collection and post-processing can be found at \cite{code}. NumPy~\cite{harris2020array}, SciPy~\cite{2020SciPy-NMeth}, Matplotlib~\cite{Hunter:2007}, tqdm~\cite{tqdm}, and PyVISA~\cite{Grecco2023} libraries are required. The code is stored in .ipynb files, they can be opened with an open-source web application Jupyter Notebook~\cite{Kluyver2016jupyter}.

\subsubsection{The assembled setup}

\begin{figure}[H]
\includegraphics[width=0.99\columnwidth]{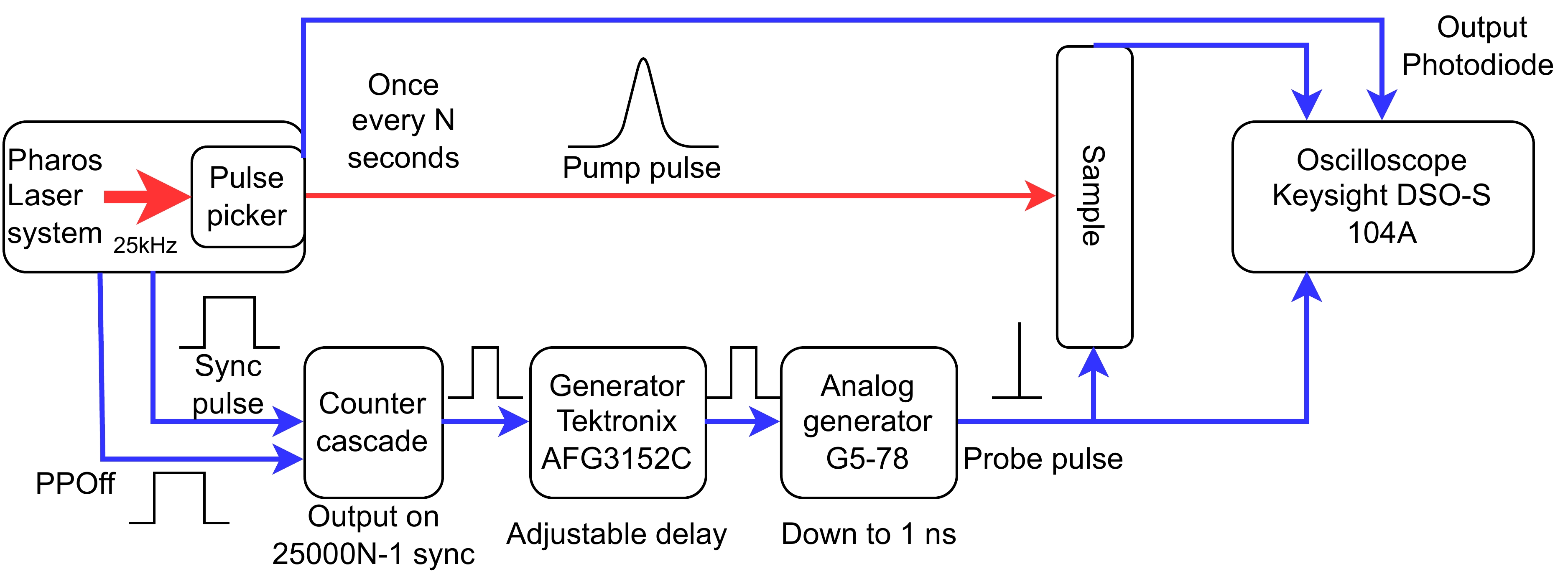}
\caption{\label{fig:exp_setup}Schematics of the experimental setup. The configuration allows down to 1 ns probe pulses.  Laser pulses are shown in red, and electric pulses are shown in blue.}
\end{figure}

The schematics of the whole experimental setup are shown in Fig. \ref{fig:exp_setup}.
In case of fast measurements (<5 ns probe pulse width), the generated pulse is used to trigger the analog pulse generator as shown in Fig. \ref{fig:exp_setup}. In case of slower measurements, the rectangular measurement pulse is generated by Tektronix AFG3152C directly, skipping the G5-78 pulse generator. The combination of two pulse generators allows both, fast probe pulses and digitally controlled delay. The duration of the probe pulse is adjusted to be sufficient for the signal to reach a steady state. The closer the impedance of the measuring circuit to 50 Ohms, the shorter pulses can be used. 

\section{\label{sec:samples}Samples}
We demonstrate the usability of our setup on two types of photodetective materials. The first one is a CdS-based photoresistor (GL5516, see datasheet: \cite{GL5516}). CdS has a direct bandgap of 2.42 eV and long-standing photoconductivity caused by depopulation of impurity levels. The dark resistance of the film is too high. Therefore, it was additionally illuminated by an incandescent lamp with a constant luminosity of 95.1 $\pm$ 0.3 lux. The luminosity was measured with a BH1750 sensor. 

The second one is a thin amorphous VO$_x$ film. The film was deposited on a glass substrate in a Plassys MEB-550S. The deposition was carried out in a flow of oxygen (8 sccm) at room temperature at a rate of 0.1 nm/s. The thickness of the films was 50 nm. In fact, this film is a thermistor with a -2.56\%/K temperature coefficient of resistivity in a wide range of temperatures (20-80 \textdegree C). Cr-Au contacts were lithographically defined and thermally evaporated using the lift-off method on top of the VO$_x$ film. Contacts have a bow tie shape with a short and wide gap in the middle (450 $\mu$m wide and 10 $\mu$m long) so the resistance of the structure was suitable for the measurement range. A photo of the structure is presented in Sec. \ref{subsec:heat}.

\section{\label{sec:results} Results and discussion}

\subsection{\label{subsec:nanosecondpulse}Demonstration of 1 ns pulses}

\begin{figure}[H]
\includegraphics[width=0.99\columnwidth]{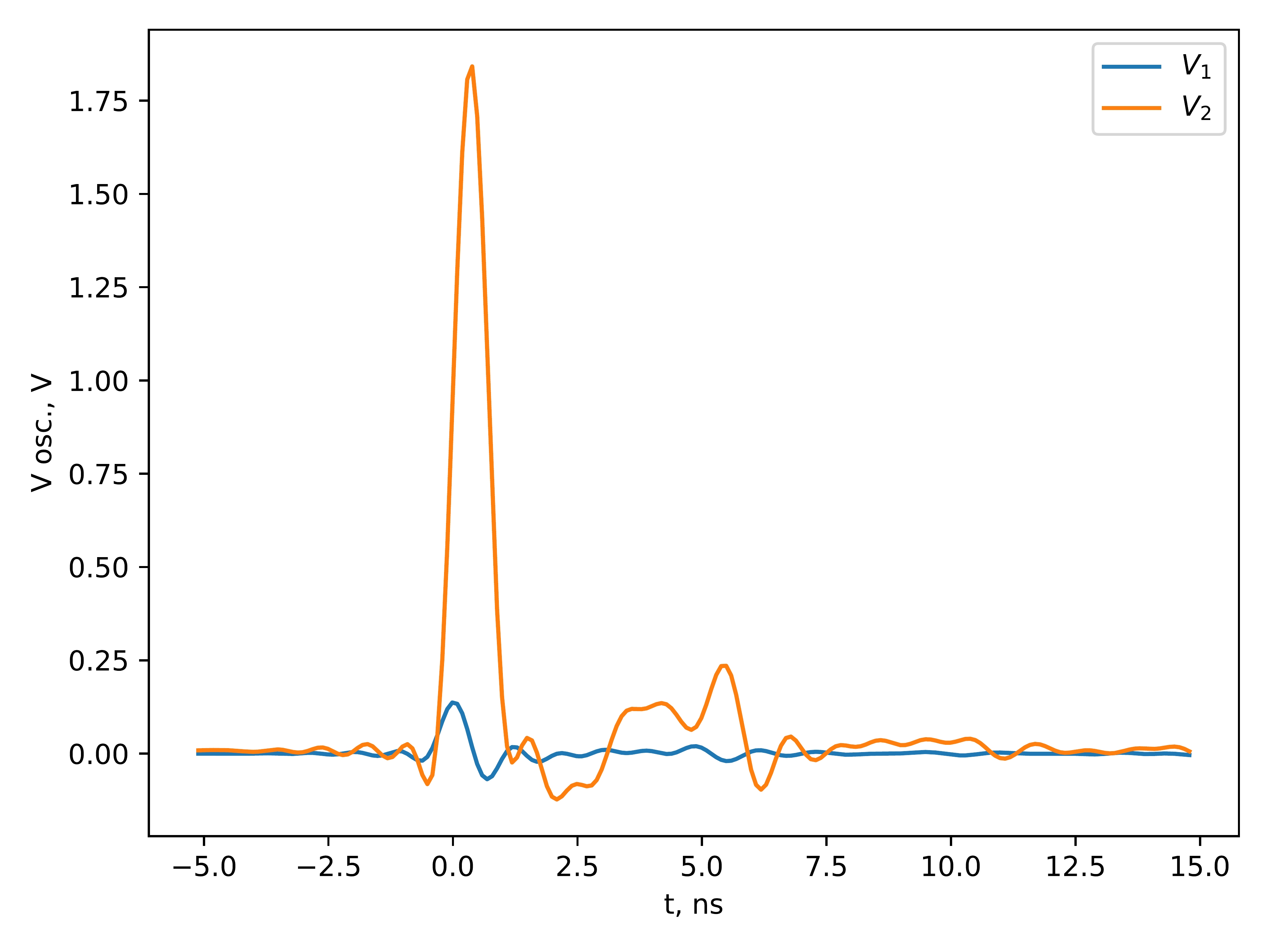}
\caption{\label{fig:nano}Oscillograms obtained from a 1 ns probe pulse. $V_1$ and $V_2$, corresponding to the first and second channels of the oscilloscope, are shown in blue and orange, respectively.}
\end{figure}

The time resolution of the method depends on the width of the probe pulse. To demonstrate that the circuit potentially allows the usage of 1 ns probe pulses, a setup with RuO$_2$ resistors is used. The laser wasn't used for the demonstration. The internal clock of the G5-78 was used to trigger the setup. Fig. \ref{fig:nano} shows oscillograms from channels 1 and 2 (in blue and orange, respectively) of the oscilloscope. $R_\text{SAMPLE}$ is 1495 Ohm and $R_M$ is 52 Ohm, measured with a Keithley 2000 multimeter. The Eq.~\ref{eq:resistance} gives 1472 Ohm. It was obtained using the integration method for determining $V_1$ and $V_2$, mentioned in Sec.~\ref{sssec:post}. o take into account the low bias voltage of the analog pulse generator we subtract the mean oscillogram value before the rectangular pulse. The amplitude of the probe pulse was 5 V, lead and trail were 500 ps. The 1.5\% disagreement shows that the probe part of the setup is capable of achieving 1 ns resolution.

\subsection{Longer pulse demonstration}

\begin{figure}[H]
\includegraphics[width=\textwidth]{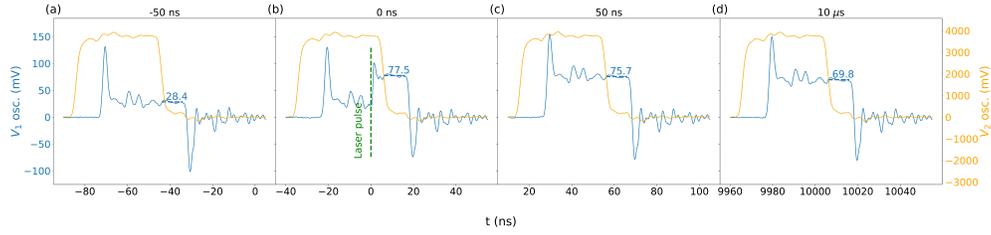}
\caption{\label{fig:long4}Oscillograms obtained at different delays for the CdS sample. Data from the first and the second channels of the oscilloscope is in blue and orange, respectively. Delays are calculated from the increase in the signal level. The stable signal level and its value are shown in blue. The laser pulse is shown in green. (a) 50 ns before the light pulse; (b) right at the light pulse; (c) 50 ns after; (d) 10 $\mu$s after. The left and right Y axes have different scales and correspond to V$_1$ and V$_2$, respectively.}
\end{figure}

Fig. \ref{fig:long4} displays oscillograms for different time delays in the CdS sample: before (a), at (b), and after the laser pulse (c, d). To take into account the 13 ns laser pulse jitter, the PHAROS output photodiode is recorded on the third channel of the oscilloscope, as mentioned in Sec.~\ref{synchproper}. The position of the photodiode voltage peak is compared to the electrical pulse position to determine the exact delay. Transient processes and voltage level changes corresponding to the laser pulse are evident. Resistance values are calculated using Eq.\ref{eq:resistance}, determining V$_1$ and V$_2$ from the first and second channel's data via one of the methods described in Sec.\ref{sssec:post}. Using 7.5V 40 ns width probe pulses for demonstration, heat estimation ($Q = U^2 / R \cdot t$) from such pulses yields approximately $10^{-15}$ to $10^{-14}$ Joules, significantly lower (9 to 8 orders of magnitude) than the laser pulse's $10^{-6}$ Joules.

\subsection{\label{subsec:longpulse}Dynamics of photoconductivity relaxation in CdS}

For CdS photoconductivity relaxation measurements, the probe pulse parameters were 25 ns width, 7 V amplitude, and 3 ns lead and trail. It's crucial to note that 7 V is the pulse generator output, not the sample voltage. Light pulses occurred every two seconds, with a matching resistor of 51 Ohm (RuO$_2$ SMD). A barium borate (BBO) crystal generated a second harmonic at $\lambda=$ 515 nm for the pump pulse. The scan duration depends on the system's relaxation time, number of shots per point and number of points per relaxation curve. For the CdS sample it takes about 2 seconds to recover. We measured 87 delay values, 20 shots each for averaging. The whole measurement process took about an hour.

\begin{figure}[H]
\includegraphics[width=0.99\columnwidth]{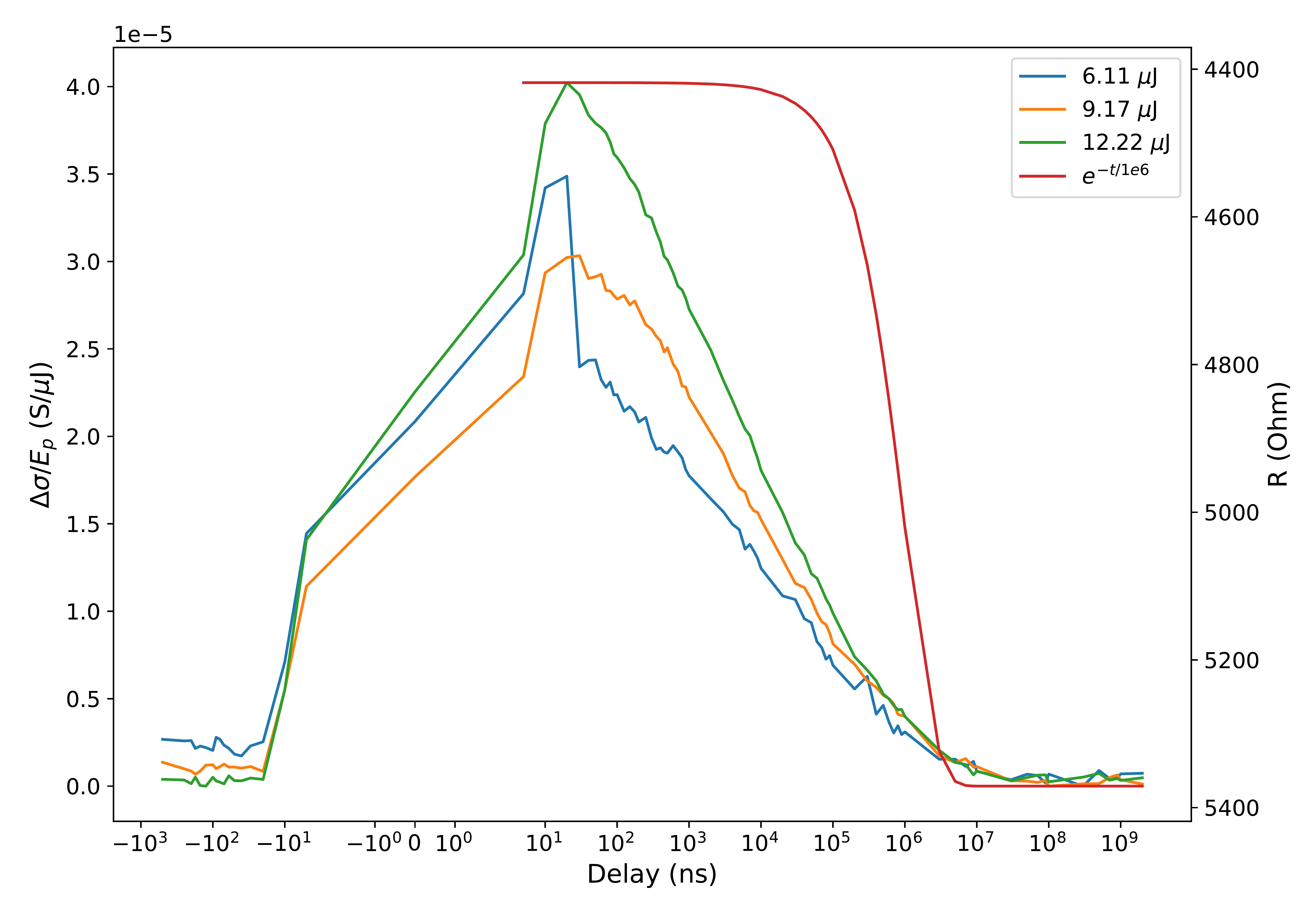}
\caption{\label{fig:relax_cds}Photoconductivity relaxation curves of a CdS photoresistor at three different pump pulse energies and an exponent for comparison (in red). The X axis has a symmetric logarithmic scale. The Y axis on the left shows photoconductance divided by the laser pulse energy. On the right it shows the corresponding resistance values.}
\end{figure}

The relaxation curve for the photoconductivity, normalized by the energy of the pulse for three pump pulse energy values is shown in Fig. \ref{fig:relax_cds}. The X-axis scale of the relaxation curve is symmetric logarithmic.

The exact position of zero delay is determined by placing the onset of the photoresistive response in the middle of the probe pulse. Because the pulse part of the oscillogram was integrated, the resistance started to drop before the zero time point. At the time point of -12.5 ns the last part of the probe pulse starts to increase as in Fig. \ref{fig:long4} (b) in the middle of the pulse.

CdS photoresistors are known to have about 100 millisecond photoconductivity relaxation time at room temperature.  In the experimental data, the relaxation begins at the sub-microsecond scale and then has rather slow dynamics. 
This behavior could be explained by many charge trap model and two contributions (fast electrons and slow holes) in photoconductivity\cite{bk:Ryvkin2012}. 
Another possible explanation is related to the states in the bandtail with a broad distribution of the relaxation times. Indeed the excitation energy (2.41 eV) is slightly less than the bandgap of CdS (2.42 eV) therefore band bottoms and band tails are effectively excited. The relaxation of itinerant carriers at the band bottom is fast, while the states in the band tails relax slowly. 

In the >100 $\mu$s region, normalized relaxation curves coincide, suggesting incomplete depopulation of the slowest traps by the pump pulse. At lower delays, distinct curves highlight the nonlinearity of light absorption. Alternatively, sub-microsecond features may result from overheating. Further studies are needed to clarify the sub-microsecond photoconductivity dynamics.

\subsection{\label{subsec:heat}Heat-induced processes}

\begin{figure}[H]
\includegraphics[width=0.99\columnwidth]{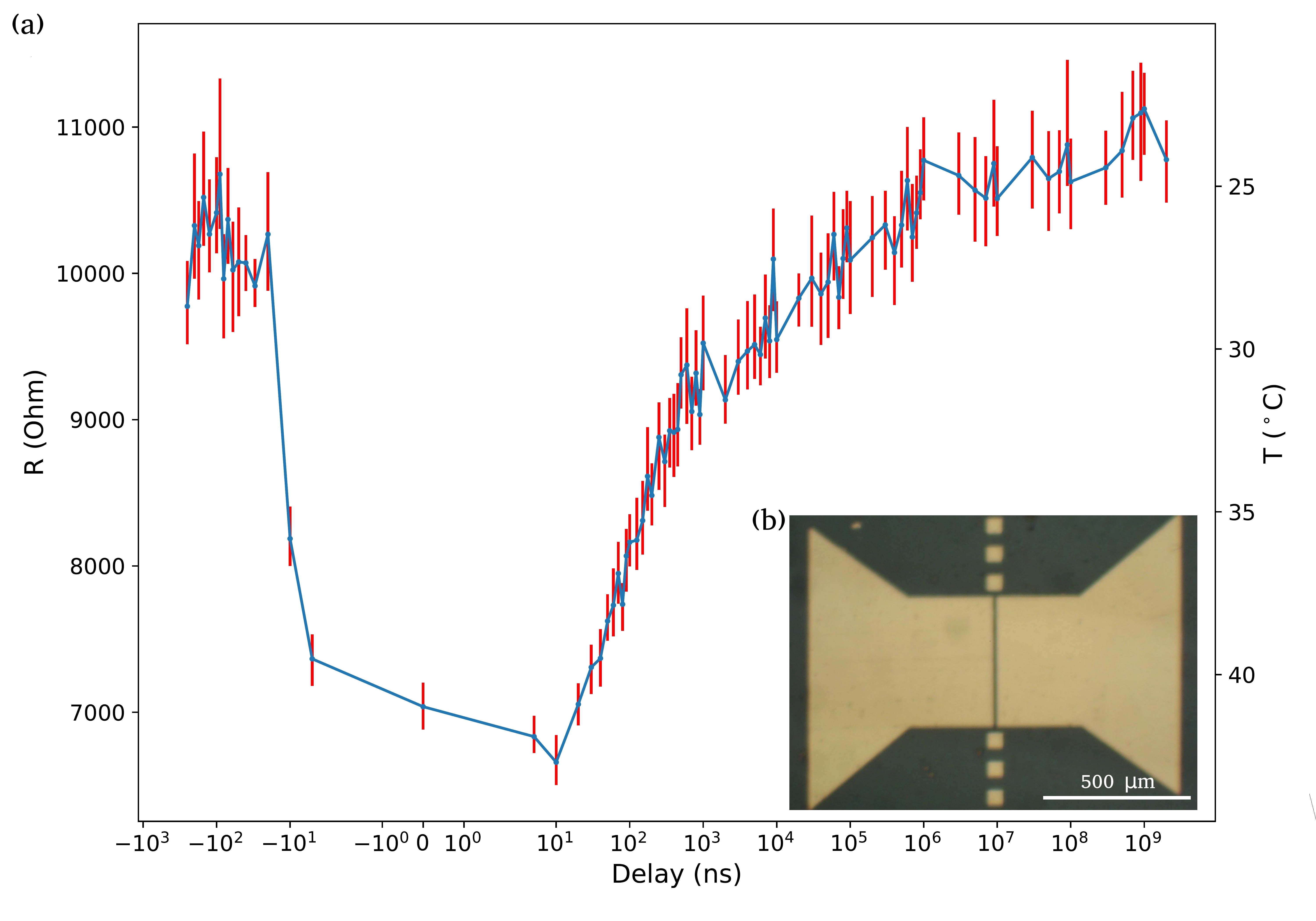}
\caption{\label{fig:relax_vox} (a) Relaxation curve for resistance and temperature of a thin amorphous VO$_x$ film on a glass substrate with errors shown in red. The X axis has a symmetric logarithmic scale. Errors were calculated using bootstrap with a 0.95 confidence interval. \label{fig:vox_photo} (b) A photo of the VO$_x$ sample on a glass substrate. Golden contacts are evaporated on top of the film. The gap between them is 10 $\mu$m long and 450 $\mu$m wide}
\end{figure}

The relaxation curve for the VO$_x$ sample is shown in Fig. \ref{fig:relax_vox}(a). In this case, the light pulse is used to heat a sample. Due to the geometry of the sample (see Fig. \ref{fig:vox_photo}(b)) and the low absorption of gold contacts and the substrate, the energy of one light pulse had to be high (110 microjoules) for registrable changes to occur. This led to the degradation of the film at the beginning of the experiment. After 5 minutes the sample was in a steady state and its resistance stopped drifting. The probe pulse parameters were 30 ns length, 7.5 V amplitude, 3 ns lead and trail. The matching resistor was a RuO$_2$ SMD 51 Ohm resistor.

No exponential or power law relaxation is observed Fig. \ref{fig:relax_vox}(a). One can see fast dynamics at low delays and slow dynamics at delays above 1 $\mu$s. We think that fast dynamics at low delays originates from heat diffusion into gold contacts with high thermal conductivity. When gold contacts are thermalized, the heat flows into the substrate slowly.

Therefore, the proposed laser-pump-resistive-probe technique in combination with a thermosensitive resistor could serve as a tool for studying heat diffusion in bulk materials and microstructures.

\subsection{\label{subsec:compare}Comparison to a DC technique}

\begin{figure}[H]
\includegraphics[width=0.99\columnwidth]{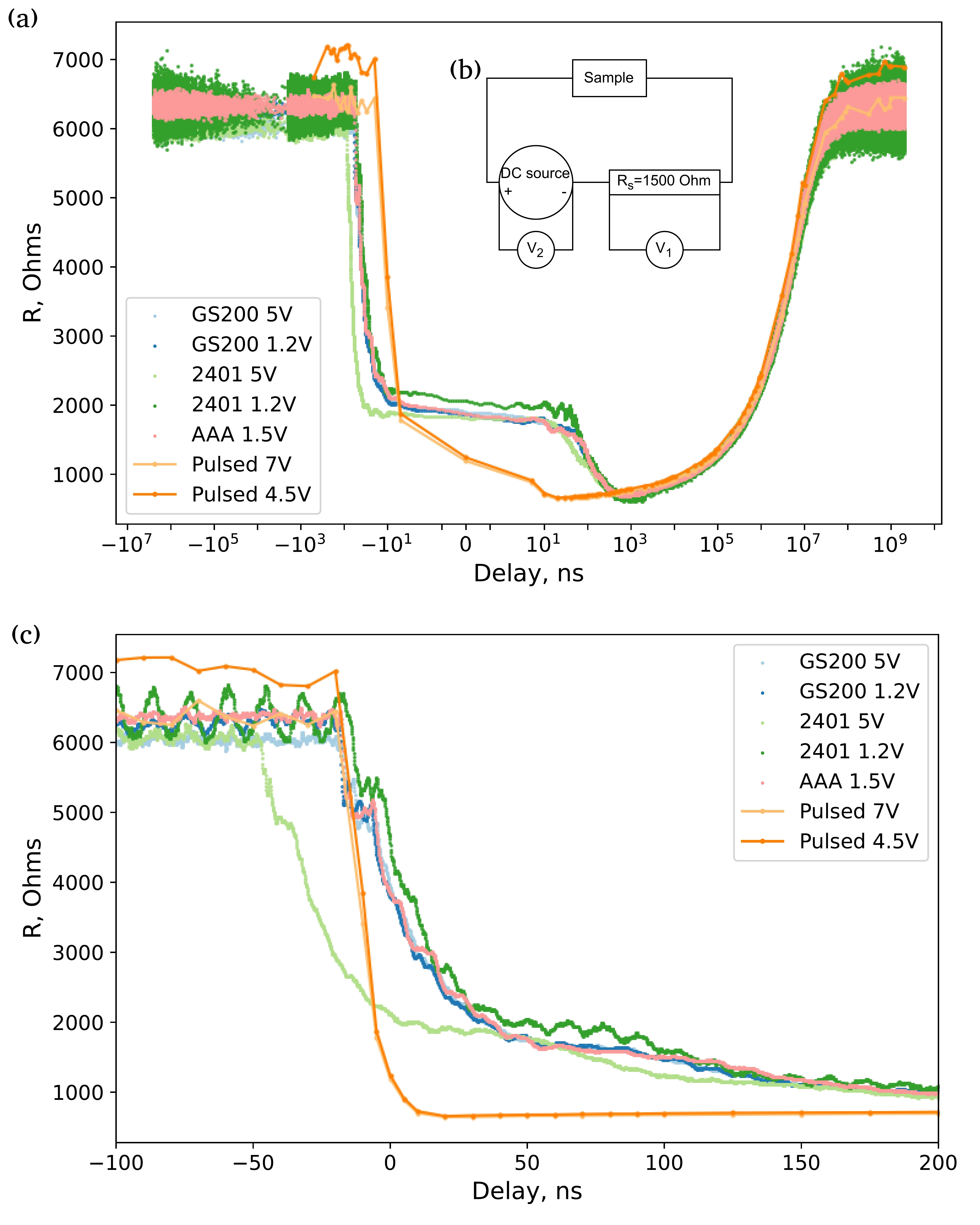}
\caption{\label{fig:dcpulse_sym} A comparison of the DC technique measurements and the proposed pulsed technique.  The DC technique trigger point was set to 50\%. (a) The X axis has a symmetric logarithmic scale. The same source with different voltages is presented as different shades of the same color; (b) \label{fig:electr_circuit_dc}Schematics of the equivalent electrical circuit for DC measurements; (c) \label{fig:dcpulse_shift} Nanosecond to hundred nanosecond range. The X axis has a linear scale. A constant 40 ns shift was applied to DC measurements.}
\end{figure}

To demonstrate the reliability of pulsed technique, we compare it with an established DC technique (e.g. Refs. \cite{CrunteanuStanescu2020,  Crunteanu}). The sample for this test is the same CdS photoresistor. The sensing resistor we use is a 1.5 kOhm precision resistor. The oscilloscope's inputs were switched to 1 MOhm. The equivalent electrical scheme of the measuring circuit is shown in Fig.~\ref{fig:electr_circuit_dc}(b).

The main advantage of the DC technique is its ease of synchronization. The oscilloscope can be triggered by the rise of the signal on its first input (a typical 50\% rise is used). Data for this method were collected in two steps. The first step was to record data from -2.5 ms to 2.5 ms with a sampling rate of 10 GSa/s. In the second step, we collected data from 0 to 2 s with a much lower sampling rate. Another advantage is that it allows capturing many more delays points. For the proposed method, one has to manually set the points to collect, as it takes a lot of time to acquire data owing to averaging. To calculate the resistance for the DC measurements, a simple formula was used (derivation in Supplement 1): 

\begin{equation}
    R_{\text{SAMPLE}} = \frac{V_2 - V_1}{V_1} \cdot R_S .
\end{equation}

We used two DC sources (YOKOGAWA GS200 and Keithley 2401) at 1.2 and 5 V, as well as a simple 1.5 V alkaline AAA battery.  For the proposed pulsed technique, two voltages were used (7 and 4.5 V). Probe pulses were 25 ns wide with a 3 ns lead and trail. The obtained curves are shown in Fig.~\ref{fig:dcpulse_sym}(a). The DC data were collected three times and averaged. Pulsed data were averaged over 17 laser pulses. The trigger was set to a typical 50\% level, so the resistance drop occurred earlier in DC technique.

In Fig.~\ref{fig:dcpulse_shift}(c) we zoom in the region from -100 ns to 200 ns, which is of particular interest. The X axis has a linear scale, and a constant 40 ns shift is applied for DC techniques, so the drop occurs at roughly the same moment.

For high delays (> 1 $\mu$s) the curves coincide. This proves the consistency of our method. For low delays (< 1 $\mu$s) the results differ strongly between the DC and pulsed measurements. Moreover, the results obtained using the DC technique differ from one power supply to another. This can be explained by the formation of an RC filter in the DC setup due to $\sim$100 pF per meter capacitance of a 50 Ohm cable.
Another reason is an improper operation of the power supply feedback loop at the nanosecond timescale. To illustrate it we show voltage drop across the DC source, V$_2$, for all power supplies up to 400 ns in Supplement 1.
Unsurprisingly, GS200 and 2401 both drop in voltage when the resistance of the circuit drops. Keithley 2401 requires much more time to stabilize (>10 $\mu$s). Thus, besides practically unlimited delay and reduced heat dissipation, another advantage of our technique in comparison with the DC measurements is the capability to measure at nanosecond scale using standard coaxial cables.

A problem of the proposed short-pulsed technique is the high scattering of the data. It can be mitigated through application of a low-pass filter, utilization of longer pulses with subsequent averaging at extended delays, or, if thermal considerations are negligible, employing the DC technique at higher delays.

At shorter delays, a single data point reflects an average across the pulse width. To improve the temporal resolution a fine adjustment of the sample geometry can be used to achieve a perfect 50 Ohm matching, as demonstrated by Stern et al.\cite{art:Stern_hallnano}. Another approach involves employing longer pulses and utilizing the oscilloscope trace, particularly if the heating effects are negligible, as illustrated in Refs. \cite{Nathawat2019, Wang2019}. The objective of the present study is to show the basic setup in an universally applicable manner.

\section{\label{sec:conclusions}Conclusions}

We developed a laser-pump-resistive-probe technique. The delay can be driven from nanoseconds to seconds. We describe the details of the method, demonstrate a nanosecond resolution and the usability of the proposed technique to study heat-induced changes and carrier relaxation.

In comparison with DC measurements and standard optical pump-probe, the proposed technique has several advantages. 
In the proposed technique, the upper delay limit is practically unlimited, in contrast to traditional optical pump-probe methods. Secondly, the thermal impact on the sample is notably lower when compared to DC techniques. Thirdly, DC measurements are prone to issues of DC source feedback and the formation of RC filters with transmission lines, whereas such challenges are absent in the pulsed technique.

The technique could be useful to study biochemical reactions, heat transfer, photoconductivity relaxation, and slow electronic transformations at a new timescale. It has also potential for several modifications. For instance, an AC probe pulse, as demonstrated in \cite{art:Dirisaglik_electrpump}, can be used to study frequency-dependent conductive responses in certain systems, such as MEMS or phase-change memory devices. Additionally, the measurement time can be improved by utilizing multiple probing pulses per pump pulse, provided the oscilloscope's memory and the system under study allow such an approach.

\begin{backmatter}

\bmsection{Funding} Ministry of Education and Science of the Russian Federation (075-15-2021-598).

\bmsection{Acknowledgments}
The authors are thankful to E. V. Tarkaeva for her contribution to the sample fabrication and to A. Yu. Klokov, K.V. Mitsen, and A.V. Varlashkin for numerous thoughtful discussions. The VO$_x$ sample fabrication has been performed at the Shared research facility at the P.N. Lebedev Physical Institute.

\bmsection{Disclosures} The authors declare no conflicts of interest.

\bmsection{Data availability} Data underlying the results presented in this paper are not publicly available at this time but may be obtained from the authors upon reasonable request.

\bmsection{Supplemental document}
See Supplement 1 for supporting content. 
\end{backmatter}

\bibliography{sample}

\end{document}